\documentclass[pra,twocolumn,aps,showpacs,preprintnumbers,amsmath,amssymb,superscriptaddress]{revtex4-1}

\usepackage{graphicx}
\usepackage{dcolumn}
\usepackage{bm}
\usepackage{amsmath}
\usepackage{hyperref}

\begin{document}

\title{Interference between independent photonic integrated devices for quantum key distribution}

\author{Henry Semenenko}
\email{henry.semenenko@bristol.ac.uk}
\affiliation{Quantum Engineering Centre for Doctoral Training, H. H. Wills Physics Laboratory \& Department of Electrical and Electronic Engineering, University of Bristol, Tyndall Avenue, BS8 1FD, UK}
\affiliation{Quantum Engineering Technology Labs, H. H. Wills Physics Laboratory \& Department of Electrical and Electronic Engineering, University of Bristol, Tyndall Avenue, Bristol, BS8 1FD, UK}

\author{Philip Sibson}
\affiliation{Quantum Engineering Technology Labs, H. H. Wills Physics Laboratory \& Department of Electrical and Electronic Engineering, University of Bristol, Tyndall Avenue, Bristol, BS8 1FD, UK}

\author{Mark G. Thompson}
\affiliation{Quantum Engineering Technology Labs, H. H. Wills Physics Laboratory \& Department of Electrical and Electronic Engineering, University of Bristol, Tyndall Avenue, Bristol, BS8 1FD, UK}

\author{Chris Erven}
\affiliation{Quantum Engineering Technology Labs, H. H. Wills Physics Laboratory \& Department of Electrical and Electronic Engineering, University of Bristol, Tyndall Avenue, Bristol, BS8 1FD, UK}

\date{\today}

\begin{abstract}
Advances in quantum computing are a rapidly growing threat towards modern cryptography. Quantum key distribution (QKD) provides long-term security without assuming the computational power of an adversary. However, inconsistencies between theory and experiment have raised questions in terms of real-world security, while large and power-hungry commercial systems have slowed wide-scale adoption. Measurement-device-independent QKD (MDI-QKD) provides a method of sharing secret keys that removes all possible detector side-channel attacks which drastically improves security claims. In this letter, we experimentally demonstrate a key step required to perform MDI-QKD with scalable integrated devices. We show Hong-Ou-Mandel interference between weak coherent states carved from two independent indium phosphide transmitters at $431$ MHz with a visibility of $46.5 \pm 0.8\%$. This work demonstrates the feasibility of using integrated devices to lower a major barrier towards adoption of QKD in metropolitan networks. 
\end{abstract}

\maketitle

\section{Introduction}
Secure communication protocols have been the focus of much academic research since the promise of quantum computing attacks against modern cryptography \cite{shor1994}. Quantum key distribution (QKD) \cite{BB84, ekert1991}, unlike many current encryption methods, does not assume the computational power of an adversary nor rely on assumed mathematically hard trap-door functions. Instead, the security of QKD is based on the laws of physics and aims to provide a long-term key exchange solution. 

In recent years, QKD has been under scrutiny from the emerging quantum hacking community who have demonstrated that real-world physical implementations do not always match the assumptions of the theoretical models \cite{Lo2014}. This can lead to malicious attacks that allow an eavesdropper to gain information about the secret key. These include side-channels \cite{Lamas-Linares:07}, where vulnerable information is leaked through uncharacterised channels, or responses to external manipulation of devices through classical means \cite{Gisin2006}. In particular, many attacks have been directed at the detectors due to their complexity and inconsistencies between theory and experiment \cite{Lydersen2010,Zhao2008, Makarov2006}.

Measurement-device-independent QKD (MDI-QKD) is a recent protocol that tackles some of the more prevalent attacks on systems by removing all detector side-channels \cite{Lo2012}. It does so by introducing a third party (Charlie) who acts as a relay to mediate detection events by announcing quantum correlations between states sent by Alice and Bob. The detection events alone do not contain any information about the secret key, so an eavesdropper cannot gain information by targeting the detectors. The protocol also lends itself to a star-shaped topology with Charlie as a shared relay. Now the detectors, and other expensive and complex equipment, can be shared between users in a metropolitan network without reducing security by introducing trusted nodes.

At the heart of the MDI-QKD protocol is Hong-Ou-Mandel (HOM) interference \cite{HOM}, a quantum phenomenon where indistinguishable single photons incident on a beam splitter interfere and bunch. HOM interference between independent sources remains challenging due to the requirement of the photons being indistinguishable in all degrees of freedom \cite{Xu2013}. It is possible to perform HOM-like interference using weak coherent states. However, due to the multi-photon nature of coherent states, the maximum visibility of a HOM dip is reduced to $50\%$ \cite{Rarity2005}. This is taken into account in MDI-QKD security proofs and does not represent a loophole \cite{Wang2013}.

\begin{figure*}[t]
	\centering
	\includegraphics[width = \linewidth]{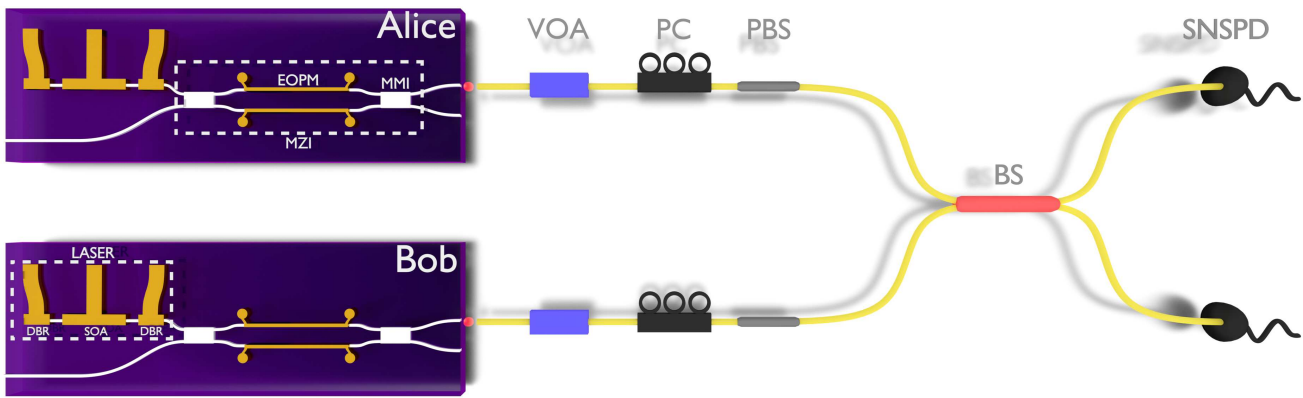}
	\caption{{\bf Experimental Setup:} Two identical 6x2 mm$^2$ indium phosphide chips were used to carve weak coherent states. Each device contains a continuous wave, on-chip, Fabry-P\'erot laser consisting of a semiconductor optical amplifier (SOA) between two distributed Bragg reflectors (DBR) which operate at $1550$ nm. The laser was carved into $120$ ps pulses using Mach-Zehnder interferometers (MZI) made from multi-mode interferometers (MMI) and electro-optic phase modulators (EOPMs). The phase modulators were pulsed from a pulse pattern generator with $2 \text{ V}_\text{pp}$ pulses.  The light was coupled off the chips using spot-size converters and collected into lensed fibres. The light was attenuated using a variable optical attenuator (VOA) and the polarisation adjusted using a polarisation controller (PC). Polarising beam splitters (PBS) were used to ensure polarisation overlap and pulses were interfered on a 50:50 fibre beam splitter (BS). Detection used superconducting nanowire single photon detectors (SNSPD) and timetags were correlated using coincidence logic.}
	\label{fig:experiment}
\end{figure*}

QKD systems have historically been bulky and expensive which has limited their practicality and has slowed their commercial adoption. Recent developments on indium phosphide (InP) photonic integrated devices have established them as a promising platform for telecommunications \cite{Meint2014} and fulfil all of the requirements to perform QKD at state-of-the-art rates \cite{Sibson2017}. The monolithic inclusion of laser sources provides an easy method of producing weak coherent states that can be used in a decoy-state QKD protocol \cite{Lo2005}. Efficient and fast phase modulation can be performed through a quantum-confined Stark effect (QCSE) with a bandwidth up to $40$ GHz \cite{Meint2014}. The possibility of mass production means that InP devices are an excellent candidate to reduce the access cost of a QKD network and allow wide adoption \cite{JeppixRoadmap}.

In this letter, we extend the application of InP devices as a QKD platform by demonstrating the required control to interfere two independent InP transmitters producing weak coherent states. We measured a HOM visibility of $46.5 \pm 0.8\%$ using pulses clocked at $431$ MHz. This visibility is comparable to other demonstrations \cite{Yuan2014, Rubenok2011, Comandar2016} with the benefit of being performed with integrated devices. Crucially, this level of interference demosntrates InP devices as a key contender as an MDI-QKD system \cite{Xu2013}. This experiment paves the way for a more practical, accessible and affordable metropolitan QKD networks.

\section{Experiment}

Two InP, monolithically fabricated transmitter devices were used to temporally modulate $120$ ps weak coherent states at a repetition rate of $431$ MHz from on-chip continuous wave lasers. The experimental setup is shown in figure \ref{fig:experiment}. The devices were independently temperature stabilised to avoid drifts in wavelength and fibre coupling. Indistinguishable pulses between the two devices was essential for maximal HOM interference. The degrees of freedom in this experiment were temporal, photon number, polarisation and wavelength.

The on-chip source was a  Fabry-P\'erot continuous wave laser made from a semiconductor optical amplifier (SOA), powered by a stable current source, and a cavity of two distributed Bragg reflectors (DBR). The laser operates in the C-band ($\sim 1550$ nm) and can be tuned in wavelength using three methods with varying precision. A broad tuning range of $>10$ nm was achieved by current injection into the DBRs, while temperature allowed for a tuning over a range of approximately $1$ nm. Ultra-fine tuning by current injection of the SOA provides steps of $80$ fm through a range of $80$ pm through effects such as heating. 

\begin{figure}[bp]
	\centering
	\includegraphics[width = \linewidth]{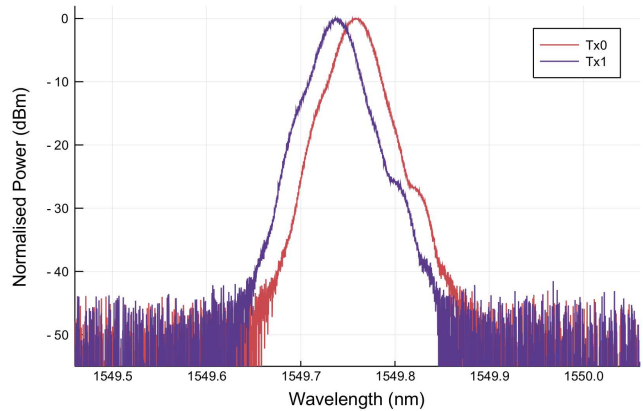}
	\caption{{\bf Laser Spectrum:} Spectral scans of the two integrated Fabry-P\'erot lasers (detuned to show each line shape) showing a $50$ dB suppression of side-bands and a FWHM of $<30$ pm (this was limited by the precision of the optical spectrum analyser).}
	\label{fig:laser}
\end{figure}

The lineshape of the Fabry-P\'erot lasers is Lorentzian \cite{Ismail16} and a spectral scan of the two transmitter lasers is shown in figure \ref{fig:laser} which shows a FWHM $<30$ pm. However, this is limited to the resolution of $30$ pm of the optical spectrum analyser (Anritsu MS9740A OSA) and the actual FWHM is expected to be much smaller. The lasers demonstrate a $>50$ dB suppression of side-bands and the small linewidth meant filtering was not required in this experiment.

The Mach-Zenhder interferometers (MZIs) are made from two $50:50$ multi-mode interferometers (MMIs) and two electro-optic modulators (EOPMs). Reverse biasing of the EOPMs creates a QCSE allowing for $10$ GHz bandwidth modulation \cite{Meint2014}. A pulse pattern generator (PPG, Keysight 81134A) provided $2 \text{ V}_\text{pp}$  pulses to carve weak coherent states using the MZIs. It is worth noting that our required voltages are much lower when compared to other modulators (e.g. lithium niobate modulators) due to the efficacy of the QCSE. 

\begin{figure}[tbp]
	\centering
	\includegraphics[width = \linewidth]{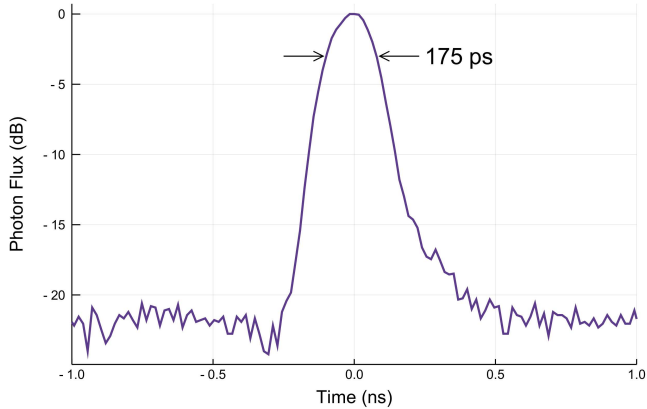}
	\caption{{\bf Pulse Carving:} Pulse shape of the $120$ ps carved lasers at $431$ MHz. The pulses have a $>20$ dB extinction ratio and FWHM on the detectors of $175$ ps, which is broadened from $120$ ps by the detector jitter ($\sim 100$ ps) and the time-tagger bin width ($32$ ps).}
	\label{fig:pulse}
\end{figure}

The FWHM of the pulses from the PPG is $120$ ps as measured on an oscilloscope (Keysight DSOX91304A) and were approximately Gaussian. A histogram of single photon detection events measuring these optical pulses is shown in figure \ref{fig:pulse}. The variation between the Gaussian electrical signal and the measured optical signal is due to signal integrity during transmission, including effects such as impedance matching. We post-select events that occur within the FWHM of the pulse to better approximate a Gaussian. The short width of the pulses in the time domain causes a Gaussian broadening in the frequency domain, which is much larger than the laser linewidth, as illustrated by the measured HOM interference.

The time of arrival of the pulses was controlled by $1$ ps precision delays in the PPG ensuring that the pulses arrived at the beam splitter at the same time. The PPGs shared a synchronised clock to ensure stability for the duration of the experiment.  A clock rate of $1.72$ GHz was used which corresponds to a $580$ ps bin width. For this experiment, a pulse was sent every 4 clock cycles ($2.32$ ns) giving a repetition rate of $431$ MHz.

Light was coupled from each transmitter device through a spot-size converter into a lensed fibre and was then passed into a digital variable optical attenuator (VOA) so that the photon number per pulse could be matched. While this experiment used external VOAs, future experiments will utilise on-chip MZIs to control attenuation. The polarisation was adjusted using a polarisation controller (PC) and passed through a polarising beam splitter (PBS) to ensure the polarisation of the two incident pulses would be maximally overlapped.  The two pulses were then interfered on a {$50$:$50$} fibre beam splitter before being sent to the detectors.

At the input to the beam splitter, the two incoming pulses can be modelled as a single frequency carved into a temporal Gaussian pulse. More explicitly, the electric field is given by
\begin{equation}
	\mathcal{E}_j^\text{in}(t) = \frac{1}{t_p\sqrt{2\pi}} \exp{\left(-\frac{t^2}{2 t_p^2}\right)} \exp{\left(i(\omega_j t + \varphi_j)\right)}
\end{equation}
where $j = 1,2$ are the beam splitter inputs. The pulse duration is given by $t_p$, the laser angular frequency is $\omega_j$ and $\varphi_j$ is the phase associated with the pulse.

After the interference through the beam splitter, the intensities of the two outputs are given by
\begin{equation}
	I^\text{out}_{1,2} = 1 \pm \exp{\left(-\frac{1}{2}t_p^2(\Delta\omega)^2\right)}\cos(\Delta\varphi)
\end{equation}
where $\pm$ distinguishes between the two output modes. We also introduce the relative phases of the pulses, $\Delta\varphi = \varphi_1 - \varphi_2$. 

From this we are able to calculate the probability of coincidence using $P(1,1) \propto \langle I_1 I_2\rangle$ \cite{Rarity2005} where we average over the relative phases, $\Delta\varphi$, of the two incident pulses. This gives the shape of the HOM dip as
\begin{equation}
	P(1,1) = 1 - V \exp{\left(-t_p^2(\Delta\omega)^2\right)}
	\label{eq:gaussian}
\end{equation}
where we introduce a visibility $V$ to take into account any distinguishability between the two input pulses. In an ideal case, we would find a maximum visibility of 50\% i.e. $V=0.5$ \cite{Rarity2005}.

\begin{figure}[bp]
	\centering
	\includegraphics[width = \linewidth]{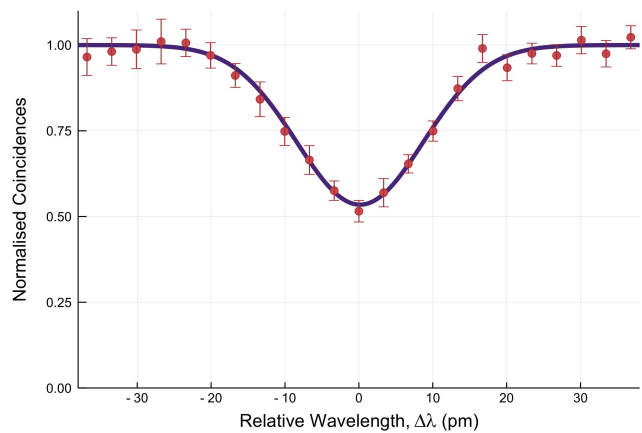}
	\caption{{\bf HOM Dip:} Hong-Ou-Mandel interference between two InP integrated devices. Using $120$ ps pulse carved from two independent on-chip lasers, we scan the wavelength of one laser to demonstrate a visibility of $46.5 \pm 0.8\%$.}
	\label{fig:HOM}
\end{figure}

The detectors used were superconducting nanowire single photon detectors (PhotonSpot) with an efficiency of ${>80\%}$, jitter of $\sim 100$ ps, dark count rate of $<500$ Hz and dead-time of $\sim 50$ ns. Detection events were collected using a PicoQuant Hydraharp and a computer was used to record coincidence events between the two detectors.

\section{Results}

Hong-Ou-Mandel interference was demonstrated between two independent InP transmitter devices by sweeping over laser wavelength through current injection of one of the lasers. Devices were initially overlapped in timing using delays on the PPG; polarisation with fibre PCs and PBSs; and photon number using digital VOAs. After this initial setup, no active feedback was required for the timing, polarisation or phase modulation demonstrating the stability of the integrated platform. The devices are temperature stabilised to $0.01^\circ$C, which provided prolonged generation of $>20$ dB pulses and stable wavelengths. Polarisation rotation in fibre was negligible during this experiment but could be stabilised in the future by actively monitoring the unused arm of the PBS.

The wavelength was varied by current injection of the laser, which also increased the laser power and was compensated for using the digital VOA. Count rates from the detectors were used as feedback to vary the attenuation and keep photon number constant. The photon number per pulse at each beam splitter input was $\approx 10^{-3}$ so as to be well away from the saturation point of the detectors and to minimise multi-photon terms that would decrease the HOM visibility.

In figure \ref{fig:HOM} we show the measured HOM interference between weak coherent states generated by the two independent InP transmitters operating at $431$ MHz. The figure plots the relative wavelength, $\Delta\lambda$, between the two lasers. As the pulses become indistinguishable, we find a bunching effect through a reduction in coincidence events between the two detectors. The singles events on both detectors remain constant throughout the sweep demonstrating that this interference is not explained through coherent classical interference where the singles events would inversely vary. A Gaussian curve (from equation (\ref{eq:gaussian})) is fitted to the data which gives a visibility of $46.5 \pm 0.8\%$. 

A misalignment in any degree of freedom between the two devices will make the pulses more distinguishable and reduce the visibility. It is for this reason that interference between independent devices remains practically challenging. The main sources of error in this experiment can be attributed to the finite extinction ratio of the pulses and mismatch between laser line and pulse shapes. 

\section{Conclusion}

In this letter, we have demonstrated the required control and precision to perform HOM interference using InP integrated photonic devices. We find a visibility of $46.5 \pm 0.8\%$ between independently generated weak coherent states at a clock rate of $431$ MHz. The interference is a vital part of the measurement process in the MDI-QKD protocol. The measured visibility is comparable to other demonstrations \cite{Yuan2014, Rubenok2011, Comandar2016} at a competitive clock rate. 

Decoy-state preparation, phase encoding and phase randomisation can be performed on a single InP device using high-speed electro-optic phase modulators. We have previously demonstrated that InP devices can perform high fidelity decoy-state BB84 states at competitive rates \cite{Sibson2017}. Together with this result, InP is shown to be a promising candidate for scalable MDI-QKD networks.

The stability and scalability of this integrated platform make it a good contender for accessible metropolitan QKD without sacrificing security by introducing trusted nodes. Access can be provided through the cheap and scalable InP platform. Expensive resources, such as detectors, optical switches and timetaggers can be shared between all users which further reduces the cost. The flexibility of the integrated photonic platform allows increased rates through wavelength division multiplexing \cite{SibsonThesis} and enables fully integrated systems through on-chip detection \cite{Akhlaghi2015} further reducing a major barrier towards widespread quantum secure communications. 

\section*{Funding} 

Engineering and Physical Sciences Research Council (EPSRC) (EP/L015730/,  EP/M013472/1, EP/L024020/1, EP/K033085/1); European Research Council (ERC) (ERC- 2014-STG 640079); QuantERA ERA-NET SQUARE project.

\section*{Acknowledgements}

During the preparation of this letter, the authors became aware of similar work by Agnesi {\it et al.} using silicon waveguide coupled III-V gain switched lasers to demonstrate HOM interference \cite{DTU-HOM}. The authors thank A. Murray for technical support, D. Aktas and G. Sinclair for useful discussions and Oclaro for the fabrication of devices through a PARADIGM project. M.G.T. acknowledges support from an ERC starter grant and EPSRC Early Career Fellowship.


\end{document}